# Pure voltage controlled magnetic nano-oscillator

Qianchang Wang[1,a], Yiheng Li[2,a], Andres Chaves[1], Joseph Schneider[1], Jin-Zhao Hu[1], Greg Carman[1,b]

a) Q. Wang and Y. Li contributed equally to this work

b) Author to whom correspondence should be addressed: carman@seas.ucla.edu

1. Department of Mechanical and Aerospace Engineering, University of California, Los Angeles, CA 90095, USA

2. Department of Modern Mechanics, University of Science and Technology of China, Hefei 230027, China

**Abstract**

Nanomagnetic oscillator is a key component for radio-frequency (RF) signal generation in many nano-scale spintronic devices. However, the actuation mechanisms of nanomagnetic oscillators are mostly current-based, which is energy inefficient at nanoscale due to Joule heating. In this study, we present a new actuation mechanism for nanomagnetic oscillator with pure voltage input using a multiferroic structure. An AC voltage with a DC bias is applied to the piezoelectric substrate, and steady perpendicular magnetic oscillation is achieved in the attached Ni disk when the frequency of the voltage matches the ferromagnetic resonance (FMR) of the Ni disk. The FMR can be tuned by simply changing the voltage bias, therefore, the oscillation frequency has a wide range. A systematic simulation study is conducted to investigate the impact of the voltage amplitude, frequency, waveform, as well as the thickness of the magnet on the magnetic oscillation. This opens new possibilities of designing energy efficient nanomagnetic oscillators using multiferroics that have large amplitude and wide frequency range.

Nanomagnetic oscillator has been used for many applications including nano-scale RF signal generator[1–3], microwave-assisted recording, nano-scale magnetic field sensor[4], and neuromorphic computing hardware[5]. In a conventional nanomagnetic oscillator, steady magnetic oscillation is achieved when the spin torque induced by the applied current cancels with the Gilbert damping.[4–10] However, the current-driven magnetic oscillation can be significantly power-consuming at nanoscale due to Joule heating. In contrast, voltage-driven magnetic oscillation can be potentially more energy efficient due to negligible Joule heating. The strain-mediated multiferroics is one of the voltage-based magnetization control mechanism. Static magnetization control by multiferroics has been demonstrated both numerically and experimentally.[11–14] There are also some work on using multiferroics for dynamic magnetization control, such as spin wave generation[15,16] and ferromagnetic resonance driven by surface acoustic wave on piezoelectric substrate[17,18]. However, there is few research on voltage-driven nanomagnetic oscillators.

One preliminary work on using strain-mediated multiferroics to drive magnetic oscillation simulates an ellipse magnet with in-plane anisotropy.[19] The magnetic oscillation is drive by a pair of off-axis electrodes on the piezoelectric substrate. The cone angle of the in-plane oscillation amplitude, however, is only 90 degree, which is limited by its mechanism of oscillating between easy and hard axes. There is lack of frequency modulation mechanism, which is one of the most important features of nanomagnetic oscillators. Therefore, a more systematic simulation is desired to develop a voltage-driven nanomagnetic oscillators with large oscillation amplitude and wide tunable frequency range.

In this study, a nano-scale Ni disk with perpendicular magnetic anisotropy (PMA) is studied as the oscillator. It is shown that the perpendicular oscillator can have large oscillation amplitude (magnetization oscillates between two easy axes $m_z = +1$ and $m_z = -1$) and wide frequency tunability.

A 3D finite element model that couples micromagnetics, electrostatics, and elastodynamics is used to simulate the strain mediated multiferroic system. Assumptions include linear elasticity, linear piezoelectricity, as well as negligible electromagnetic radiation, thermal fluctuations and magnetocrystalline anisotropy. The precessional magnetic dynamics are governed by the Landau-Lifshitz-Gilbert (LLG) equation:

$$\frac{\partial \bm{m}}{\partial t} = -\mu_0 \gamma (\bm{m} \times \bm{H}_{eff}) + \alpha \left(\bm{m} \times \frac{\partial \bm{m}}{\partial t}\right) \quad (1)$$

where $\bm{m}$ is the normalized magnetization, $\mu_0$ is the vacuum permittivity, $\gamma$ is the gyromagnetic ratio and $\alpha$ is the Gilbert damping parameter. $\bm{H}_{eff}$ is the effective magnetic field defined by $\bm{H}_{eff} = \bm{H}_{ex} + \bm{H}_{Demag} + \bm{H}_{PMA} + \bm{H}_{ME}$, where $\bm{H}_{ex}$ is the exchange field, $\bm{H}_{Demag}$ the demagnetization field, $\bm{H}_{PMA}$ the effective PMA field, and $\bm{H}_{ME}$ the magnetoelastic field by strain. The PMA field is calculated by $\bm{H}_{PMA} = -2K_{PMA} m_z \hat{\bm{z}}/(\mu_0 M_S)$.[14,20] Assuming the PMA origins from interfacial effect, and it is calculated as $K_{PMA} = K_i/t$, where $t$ is the thickness of the magnetic thin film, and $K_i = 2.6 \times 10^{-4}$ J/m² for Ni. The $\bm{H}_{ME}$ field is calculated as[21]:

$$\bm{H}_{ME}(\bm{m}, \bm{\varepsilon}) = -\frac{1}{\mu_0 M_S} \frac{\partial}{\partial \bm{m}} \{B_1 [\varepsilon_{xx}\left(m_x^2 - \frac{1}{3}\right) + \varepsilon_{yy}\left(m_y^2 - \frac{1}{3}\right) \\ + \varepsilon_{zz}\left(m_z^2 - \frac{1}{3}\right)] + 2B_2(\varepsilon_{xy} m_x m_y + \varepsilon_{yz} m_y m_z + \varepsilon_{zx} m_z m_x)\} \quad (2)$$

where $m_x$, $m_y$ and $m_z$ are components of normalized magnetization along $x$, $y$ and $z$ axis, $B_1$ and $B_2$ are first and second order magnetoelastic coupling coefficients. $B_1$ and $B_2$ are defined by: $B_1 = B_2 = \frac{3E\lambda_S}{2(1+\nu)}$, where E is the Young's modulus and $\lambda_S$ is the saturation magnetostriction coefficient of the magnetic material. The Ni material parameters used in the analysis are: $\alpha = 0.038$, $M_s = 4.8 \times 10^5$ A/m, exchange stiffness $A_{ex} = 1.05 \times 10^{-11}$ J/m (used in $\bm{H}_{ex}$), and $\lambda_s = -34$ ppm, Young's modulus $E = 180$ GPa, density $\rho = 8900$ kg/m³, and Poisson's ratio $\nu = 0.31$.[22–25]

In equation 2 for $\bm{H}_{ME}$, $\bm{\varepsilon}$ is the total strain consisting of two parts: $\bm{\varepsilon} = \bm{\varepsilon}_p + \bm{\varepsilon}^m$, where $\bm{\varepsilon}_p$ is the piezostrain, and $\varepsilon_{ij}^m = 1.5 \lambda_s (m_i m_j - \delta_{ij}/3)$ is the strain contribution due to isotropic magnetostriction, $\delta_{ij}$ is Kronecker delta function[21]. The piezostrain $\bm{\varepsilon}_p$ is determined using the linear piezoelectric constitution equations:

$$\bm{\varepsilon}_p = s_E : \bm{\sigma} + d^t \cdot \bm{E} \quad (3)$$

$$\boldsymbol{D} = d{:}\boldsymbol{\sigma} + e_\sigma \cdot \boldsymbol{E} \tag{4}$$

where $\boldsymbol{\sigma}$ is stress, $\boldsymbol{D}$ is electric displacement, $\boldsymbol{E}$ is electric field, $s_E$ is the piezoelectric compliance matrix under constant electric field, $d$ and $d^t$ are the piezoelectric coupling matrix and its transpose, and $e_\sigma$ is electric permittivity matrix measured under constant stress. More details about the simulation setup can be found in the publications by Liang et al.[12,26]

Figure 1(a) illustrates the simulated multiferroic structure. A PZT-5H (simplified as PZT below) substrate is used as the piezoelectric material with lateral size of 1500 nm × 1500 nm and 800 nm thickness. The PZT's top surface is mechanically free, and its bottom surface is fixed (i.e. mechanically clamped on a thick substrate) and low-reflecting boundary conditions are applied to the four lateral sides. A Nickel magnetic disk with a diameter of 50 nm and a height of 2 nm is perfectly adhered in the center of the PZT top surface. Two 50 nm × 50 nm square electrodes are placed symmetrically adjacent to the Ni disk along $y$ axis. The edge-to-edge distance between the electrode and the magnetic disk is 20 nm. Voltage pulses are always applied to or removed from the two electrodes simultaneously, while the bottom surface of PZT is electrically grounded.

Figure 1(b) and (c) provide the results of magnetic precession for the Ni disk without and with applied voltage, respectively. In both figures, the magnetization is released from a canted direction $\boldsymbol{m} = (0, 1, 1)/\sqrt{2}$. Fig. 1(b) illustrates the 3D trajectory for a 5-ns magnetic precession without applied voltage. In this result the magnetic anisotropy is dominated by PMA with the effective field $\boldsymbol{H}_{eff}$ along the $z$ direction. In contrast, Fig. 1(c) presents the results with an applied 1.8 V to the electrodes. The voltage induces a compressive strain along the $y$ axis and a tensile strain along $x$ axis between the two top electrodes. This strain combined with the negative magnetostrictive of Ni produces a dominating magnetoelastic field along the $y$ axis. Without applied voltage, there are two stable states $m_z = +1$ and $m_z = -1$. Applying voltage brings the magnetization to an intermediate state, i.e., in-plane. By accurately timing the voltage application, it is possible to oscillate the magnetization perpendicularly between $m_z = +1$ and $m_z = -1$.

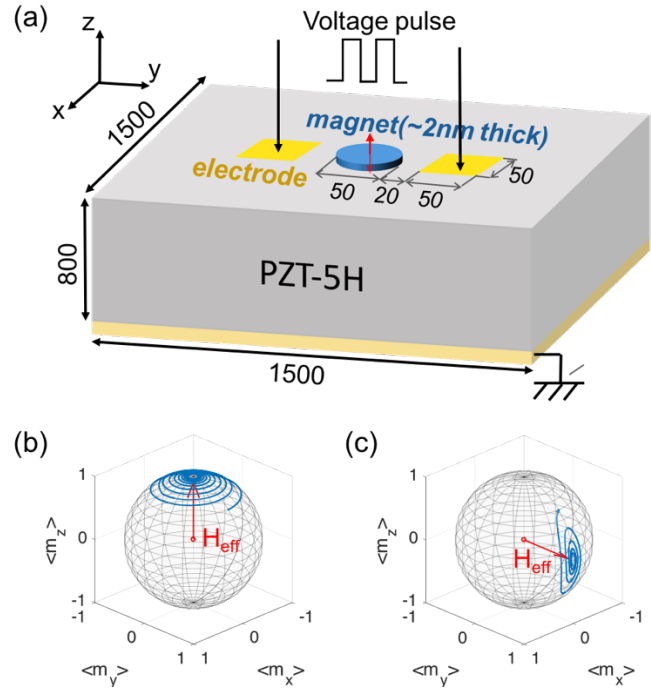

Fig. 1. (a) 3D illustration of the simulated structure (unit: nm). (b) Trajectory of magnetic precession without voltage applied. (c) Trajectory of magnetic precession when +1.8 V is applied to the top electrodes.

Figure 2(a)-(d) present simulation results for four different voltage input conditions. In the figures, the blue dashed line represents the applied voltage while the solid black line represents the Ni disk's volume-averaged perpendicular magnetization $m_z$ as a function of time. All four voltage inputs are square waves with minimum value 0 and maximum value $V_0$. The initial voltage at t = 0 is $V_0/2$ and ramps towards $V_0$. All ramps occur in 0.1 ns for 1.1 GHz voltage and the portion of the ramp within a period is kept the same for voltages applied at other frequencies. This is to eliminate the effect of voltage ramp on the magnetic dynamics.

Figure 2(a)(b)(c) show the results for 1.8V amplitudes at 0.8 GHz, 1.1 GHz and 1.6 GHz, respectively. The $m_z$ temporal response for Fig. 2(a) and 2(c) show disordered magnetic oscillation as contrasted with Fig. 2(b) which shows a steady-state magnetic oscillation, i.e., the amplitude of 180° reorientation between $m_z$ =1 to -1 kept almost unchanged in several periods. This is explained as follows. The magnetic oscillation for each case consists of two stages. The first stage is 180° perpendicular switching from mz = +1 and mz = -1 as the voltage is initially turned on and reaches 1.8 V. This stage requires accurate timing, i.e., the frequency of applied voltage should match ferromagnetic resonance (FMR) and this is only satisfied in the case shown in Fig. 2(b). Turning off the voltage too late or too early leads to either over- or under-shooting the $m_z = -1$, hence disrupt the magnetic oscillation, as shown in Fig. 2(a) and (c), respectively. The second stage is small perturbation near $m_z = \pm1$ during voltage-off half period. This stage does not require accurate timing and the voltage is designed to be symmetric for simplicity, i.e., the voltage-off and voltage-on periods have the same length in Fig. 2.

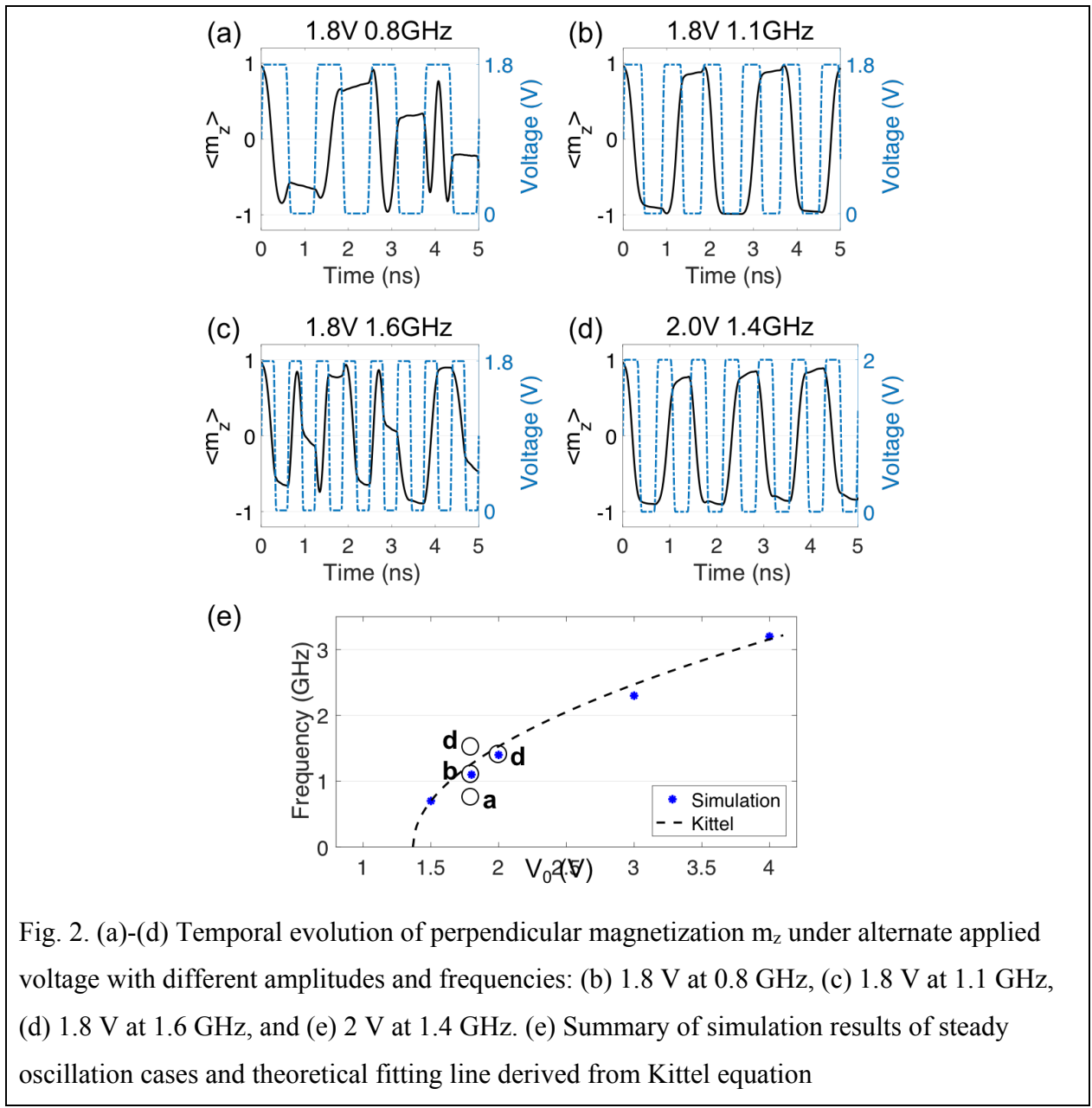

Fig. 2. (a)-(d) Temporal evolution of perpendicular magnetization $m_z$ under alternate applied voltage with different amplitudes and frequencies: (b) 1.8 V at 0.8 GHz, (c) 1.8 V at 1.1 GHz, (d) 1.8 V at 1.6 GHz, and (e) 2 V at 1.4 GHz. (e) Summary of simulation results of steady oscillation cases and theoretical fitting line derived from Kittel equation

The oscillation frequency is tunable as shown by another oscillation case with applied voltage 2 V at 1.4 GHz in Fig. 2(d). This is because increasing the voltage amplitude increases the FMR of the magnetic disk. It can be inferred that the upper limit of the frequency of the magnetic oscillator is mainly restricted by the breakdown field of the PZT substrate, and the lower limit is determined by the minimum voltage required to overcome PMA and initiate the oscillation.

To better understand the mechanism of frequency modulation, Fig. 2(e) summarizes five steady oscillation cases (marked by blue stars) with a theoretical fitting curve derived from Kittel equation. The four cases discussed in Fig. 2(a)-(d) are also marked on Fig. 2(e) by black circles. There is a good agreement between the Kittel equation and the steady oscillation cases. This confirms that the key of tuning frequency of the magnetic oscillator is shifting FMR of the magnetic disk by applied voltage.

The derivation of the theoretical fitting line in Fig. 2(e) is shown below. For a thin disk as simulated in this work, the demagnetization factors are approximately $N_x = N_y = 1$, $N_z = 0$,[27] and the coordinate is defined in Fig. 1. Then the Kittel equation is simplified as:[28]

$$f = \frac{\gamma \mu_0}{2\pi} \sqrt{H_{eff}(H_{eff} + M_S)} \quad (5)$$

Assume the strain along $y$ axis $\varepsilon_{yy}$ as the main contributing component to the magnetoelastic field because other strain components are either tensile or negligibly small, the magnetoelastic can be expressed as:

$$H_{me} = -\frac{2}{\mu_0 M_S} B_1 m_y \varepsilon_{yy} \quad (6)$$

Then the total effective field can be written as:

$$H_{eff} = -\frac{2}{\mu_0 M_S} B_1 m_y (\varepsilon_{yy} + 2 \times 10^{-3}) \quad (7)$$

Here the PMA effect is taken into consideration as a preset 2000 ppm strain given the fact that $\varepsilon_{yy} \approx -2000\ ppm$ is the minimum required strain to overcome PMA. To further simplify the calculation, we take $m_y = 0.2$ instead of a temporally variant value.

Plugging equation (7) into (5) results in an equation of the frequency as a function of $\varepsilon_{yy}$. Then $\varepsilon_{yy}$ and voltage amplitude are related by a linear equation: $\varepsilon_{yy} = -1463 \times V_0\ (ppm)$. This is obtained from a stationary simulation and more details can be found in section A in Supplemental Material. The analytical expression of frequency as a function of voltage amplitude is drawn as the dashed line in Fig. 2(e).

Except for changing voltage amplitude, another way to shift the FMR of the magnetic disk is by changing its geometry. Fig. 3 compares the results of the multiferroic magnetic oscillators with 2 nm and 1.8 nm thicknesses. Fig. 3(a) shows that the 4 V at 3.2 GHz can excite steady magnetic oscillation, however, the same voltage application does not work when the thickness is decreased to 1.8 nm, as shown in Fig. 3(b). As shown in Fig. 3(c), the steady oscillation occurs again for the 1.8 nm thick disk when the applied voltage is increased to 4.3 V. The explanation is as follows. The PMA is inversely proportional to the magnet's thickness, so the 1.8 nm oscillator has stronger PMA hence requires higher voltage to overcome the PMA. This means the FMR curve for the 1.8 nm oscillator will shift towards right compared to the 2 nm oscillator (shown in Fig. 2(e)), as the intersection point of the FMR curve on $x$ axis corresponds to the minimum voltage required to overcome PMA. Therefore, the 1.8 nm oscillator requires higher voltage to achieve a steady magnetic oscillation at the same frequency.

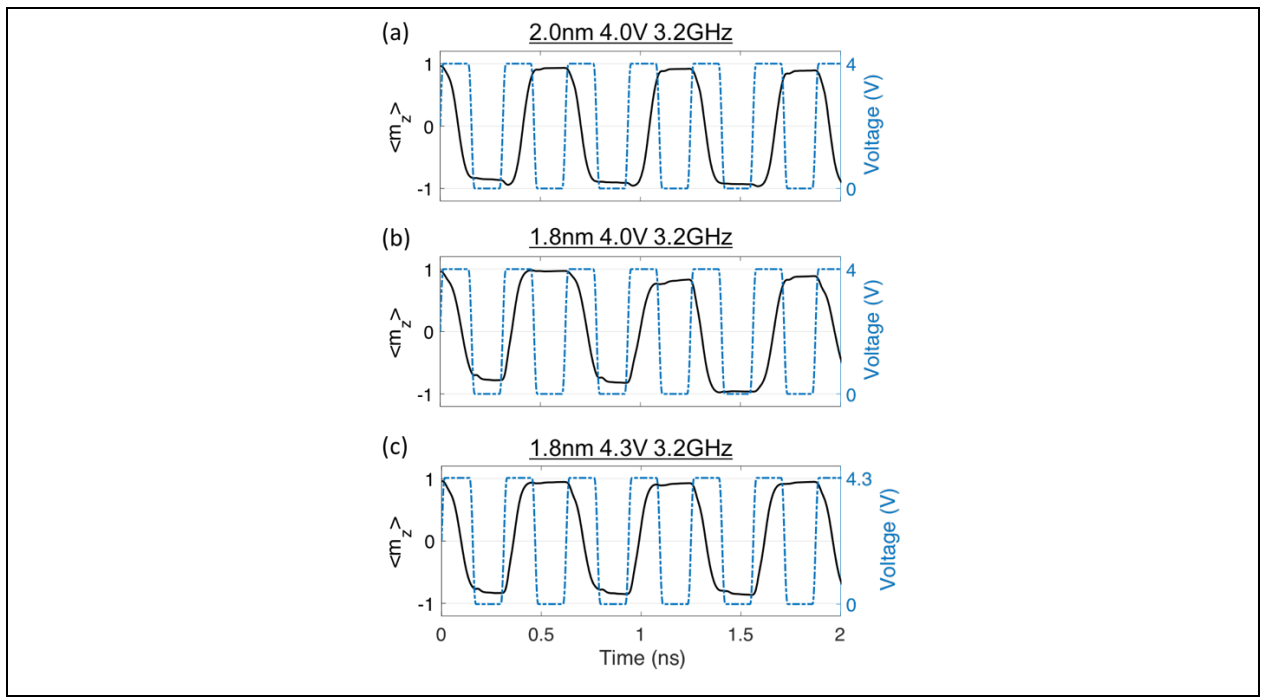

Fig. 3. Temporal evolution of perpendicular magnetization $m_z$ for magnets with different thicknesses. (a) 2nm-thick magnet with 4 V applied voltage at 3.2 GHz. (b) 1.8nm-thick magnet with 4 V applied voltage at 3.2 GHz. (c) 1.8nm-thick magnet with 4.3 V applied voltage at 3.2 GHz.

All the cases discussed above have symmetric applied voltage profile, i.e., the voltage-on and voltage-off have the same temporal length within each period of voltage profile. As discussed previously, the voltage-on stage requires accurate timing and should match the FMR. In contrast, the length of voltage-off stage has more flexibility and can be tuned to achieve an arbitrary overall oscillation frequency. Fig. 4(a) and (b) compare the 1.8 V applied voltage at 0.55 GHz with symmetric and asymmetric profile, respectively. No steady magnetic oscillation is achieved in Fig. 4(a) because the frequency of the voltage does not match the FMR of the magnetic disk, which is 1.1 GHz at 1.8 V. In contrast, the voltage-on portion in Fig. 4(b) is designed to matches the FMR of 1.1 GHz, but the voltage-off portion is purposely extended. Consequently, a steady magnetic oscillation with an overall much lower frequency (i.e., 0.55 GHz) is achieved by using the asymmetric voltage profile.

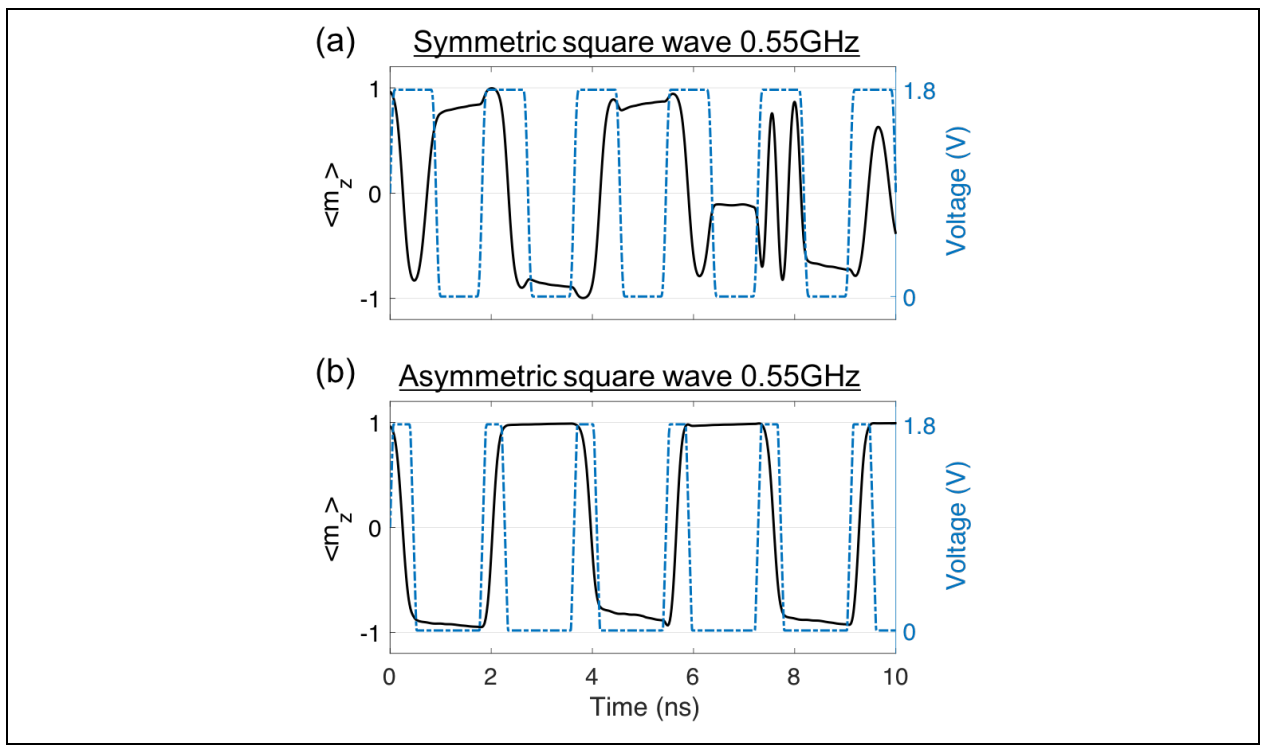

Fig. 4. Temporal evolution of perpendicular magnetization m$_z$ under alternate applied voltage with (a) symmetric square wave at 0.55 GHz and (b) asymmetric square wave at 0.55 GHz.

Figure 5 examines the impact of voltage waveforms in a multiferroic nanomagnetic oscillator. Fig. 5(a) has the simple square wave with 1.8 V amplitude and steady oscillation is achieved when the voltage has 1.1 GHz frequency. As shown in Fig. 5(b), simply changing the square wave to a sinusoidal wave with the same amplitude $V = 0.9 + 0.9\sin(2\pi f_0 t)$ ($f_0 = 1.1$ GHz) does not result in a similar magnetic response. Instead, we take the zeroth- and first-order components of the Fourier series expansion of the square wave to build the wave $V = 0.9 + \frac{4}{\pi} \times 0.9\sin(2\pi f_0 t)$, and a steady magnetic oscillation is achieved as shown in Fig. 5(c). In other words, the sinusoidal wave built in this way is equivalent to the square wave in exciting magnetic oscillation. Since sinusoidal waves are easier to generate by source meters than square waves, this provides a potentially easier way to achieve the purely voltage driven magnetic oscillations in future devices.

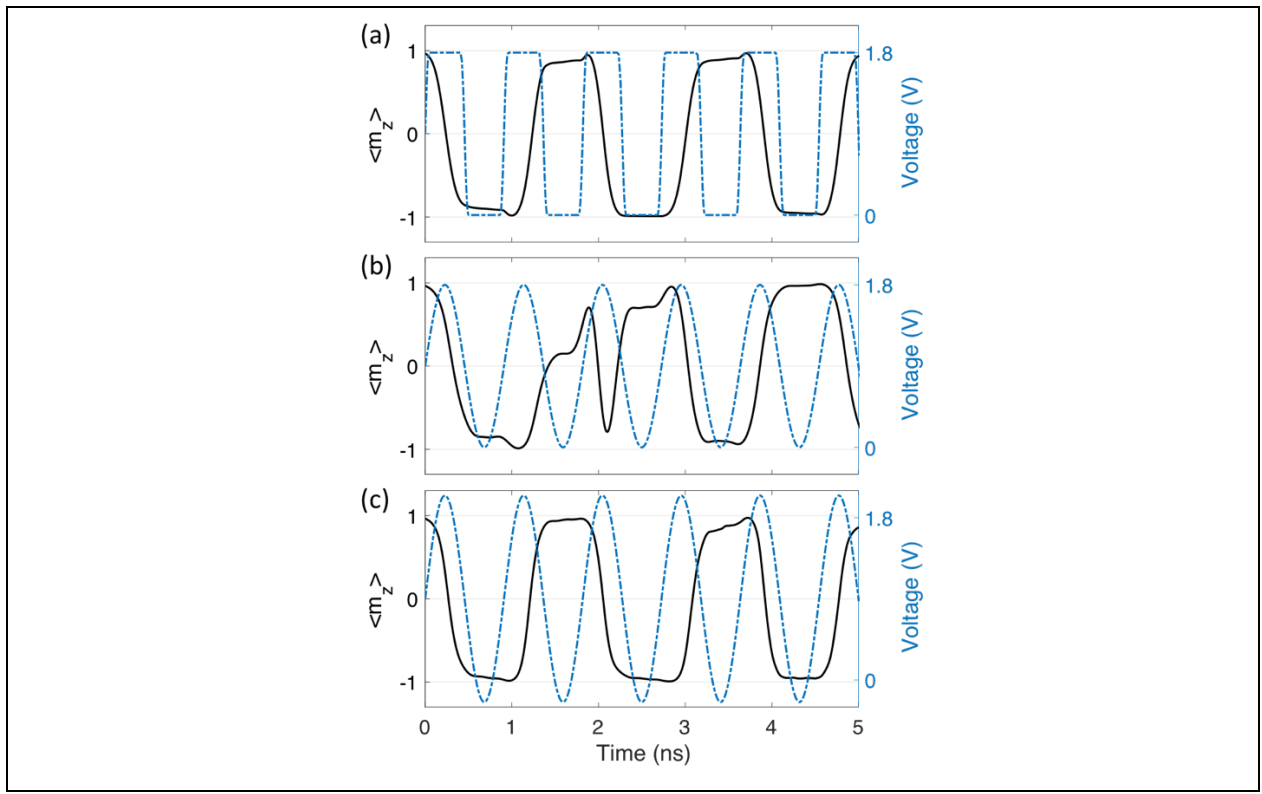

Fig. 5. Temporal evolution of perpendicular magnetization $m_z$ for applied voltage in different waveforms. (a) Square wave with 1.8 V amplitude at 1.1 GHz frequency. (b) Sinusoidal wave $V = 0.9 + 0.9\sin(2\pi f_0 t)$. (c) Sinusoidal wave $V = 0.9 + \frac{4}{\pi} \times 0.9\sin(2\pi f_0 t)$, where $f_0 = 1.1$ GHz.

In conclusion, a new magnetic oscillator mechanism is proposed and an alternate voltage applied to the piezoelectric substrate can excite steady magnetic oscillation. The oscillation frequency can be tuned by changing the FMR of the magnet, either by changing the amplitude of the alternate voltage or by changing the thickness of the magnet. The frequency range achieved in this study is from 275 MHz to 1.6 GHz (note the magnetic oscillation frequency is half of the voltage frequency). Using an asymmetric voltage profile adds more tunability the system and further extends the lower bound of the oscillation frequency. A simplified analytical equation is derived to link the oscillation frequency and the voltage amplitude. This helps understand the working principle of the purely voltage driven magnetic oscillator and guide future design of the oscillator with specified frequency range.